\documentclass[epj]{svjour}
\usepackage{graphicx}
\usepackage{amsmath,amssymb}
\usepackage{changes}
\usepackage{color}

\begin{document}
\sloppy

\title{Extinction phase transitions in a model of ecological and evolutionary dynamics}

\author{Hatem Barghathi\inst{1,2}\and Skye Tackkett\inst{1}\and Thomas Vojta\inst{1}}

\institute{Department of Physics, Missouri University of Science and Technology, Rolla, Missouri 65409, USA \and
           Department of Physics, University of Vermont, Burlington, Vermont 05405, USA}
	
\date{Received:  / Revised version: }
	
\abstract{
We study the non-equilibrium phase transition between survival and extinction of spatially
extended biological populations using an agent-based model. We especially focus on the
effects of global temporal fluctuations of the environmental conditions, i.e., temporal disorder.
Using large-scale Monte-Carlo simulations of up to $3\times 10^7$ organisms and $10^5$ generations,
we find the extinction transition in time-independent environments to be in the
well-known directed percolation universality class. In contrast, temporal disorder leads to a
highly unusual extinction transition characterized by  logarithmically slow population decay and enormous
fluctuations even for large populations. The simulations provide strong evidence for
this transition to be of exotic infinite-noise type, as recently predicted by
a renormalization group theory. The transition is accompanied by temporal Griffiths phases featuring
a power-law dependence of the life time on the population size.
}

	
\maketitle
	
\section{Introduction}\label{sec:intro}

The scientific study of population growth and extinction has a long history.
It dates back, at least, to the efforts of Euler and Bernoulli in the 18th century
\cite{Euler1748,Bernoulli1760}.
Early work was based on simple deterministic equations in which any spatial dependence
was averaged out. Later, stochastic versions of these models were also considered.
Recently, attention has turned to approaches that provide more realistic descriptions
by incorporating fluctuations in space and time as well as features such as
heterogeneity and mobility (see, e.g., Refs.\ \cite{Bartlett61,Britton10,OvaskainenMeerson10}
and references therein).

The behavior of a biological population close to the transition between survival and
extinction is of great conceptual and practical importance. On the one hand,
extinction transitions are prime examples of non-equilibrium phase transitions between
active (fluctuating) and inactive (absorbing) states, a topic of considerable current interest
in statistical physics \cite{MarroDickman99,Hinrichsen00,Odor04,HenkelHinrichsenLuebeck_book08}.
On the other hand, efforts to predict and perhaps
even prevent the extinction of biological species on earth require a thorough understanding of
the underlying mechanisms. The same also holds for the opposite problem, viz., efforts
to predict, control and eradicate epidemics.

(Random) temporal fluctuations of the environment play a crucial role
for the behavior of a biological population close to extinction.  In contrast to
intrinsic demographic noise, environmental noise causes strong population fluctuations
even for large populations which makes extinction easier. As a result, the mean time to
extinction for uncorrelated environmental noise only grows as a power of the population
size rather than exponentially as it would for time-independent environments
\cite{Leigh81,Lande93}. Recent
activities have also analyzed the effects of noise correlations in time on these results (for a
recent review, see, e.g., Ref.\ \cite{OvaskainenMeerson10}).

Most of the above work focused on space-independent (single-variable or mean-field)
models of population dynamics. Significantly less is known about the effects of temporal
environmental fluctuations on the dynamics of spatially extended
populations and, in particular, on their extinction transition.
Kinzel \cite{Kinzel85} established a general criterion for the stability of a
non-equilibrium phase transition against such temporal disorder.
Jensen \cite{Jensen96,Jensen05} studied directed bond percolation with temporal disorder and
reported that the extinction transition is characterized nonuniversal critical
exponents that change continuously with disorder strength.
Vazquez et al.
\cite{VBLM11} demonstrated that the power-law dependence between population size
and mean time to extinction also holds for spatially extended systems in some parameter region
around the extinction transition (which they called the temporal Griffiths phase).

More recently,
Vojta and Hoyos \cite{VojtaHoyos15} used renormalization group arguments to predict a highly unconventional
scenario, dubbed infinite-noise criticality, for the extinction transition in the
presence of temporal environmental fluctuations. This prediction was confirmed for the one-dimensional and two-dimensional contact process by extensive Monte-Carlo simulations  \cite{BarghathiVojtaHoyos16}.

Here, we study the transition between survival and extinction of a spatially extended biological population
by performing large-scale Monte-Carlo simulations of an agent-based off-lattice model.
In the absence of temporal environmental fluctuations, we observe a conventional extinction transition in the
well-known directed percolation universality class of non-equilibrium statistical physics.
Environmental fluctuations are found to destabilize the directed percolation behavior. The resulting extinction transition
is characterized by logarithmically slow population decay and enormous
fluctuations even for large populations. Numerical data for the time-dependence of the average population size,
the spreading of the population from a single organism, as well as the mean time to extinction
are well-described within the theory of infinite-noise critical behavior \cite{VojtaHoyos15}.

Our paper is organized as follows. In Sec.\ \ref{sec:model}, we introduce our model and the
Monte-Carlo simulations. Section \ref{sec:nodisorder} reports on the extinction transition
in time-independent environments, while section \ref{sec:disorder} is devoted to the case
of temporally fluctuating environments. In the concluding Sec.\ \ref{sec:conclusions}, we discuss
how general these findings are, we compare them to earlier simulations,
and we discuss generalizations to correlated disorder as well as spatial inhomogeneities.

\section{Model and simulations}\label{sec:model}
\subsection{Definition of the model}\label{subsec:model}

The model was originally conceived \cite{DeesBahar10,SKMB13} as a model
of evolutionary dynamics in phenotype space but it can be interpreted as
a model for population dynamics in real space as well. Organisms reside
in a continuous two-dimensional space of size $L\times L$. In the evolutionary
interpretation, the two coordinates correspond to two independent phenotype
characteristics (in arbitrary units) while they simply represent the real
space position of the organism for population dynamics.

\begin{figure}
\centerline{\includegraphics[width=8.1cm]{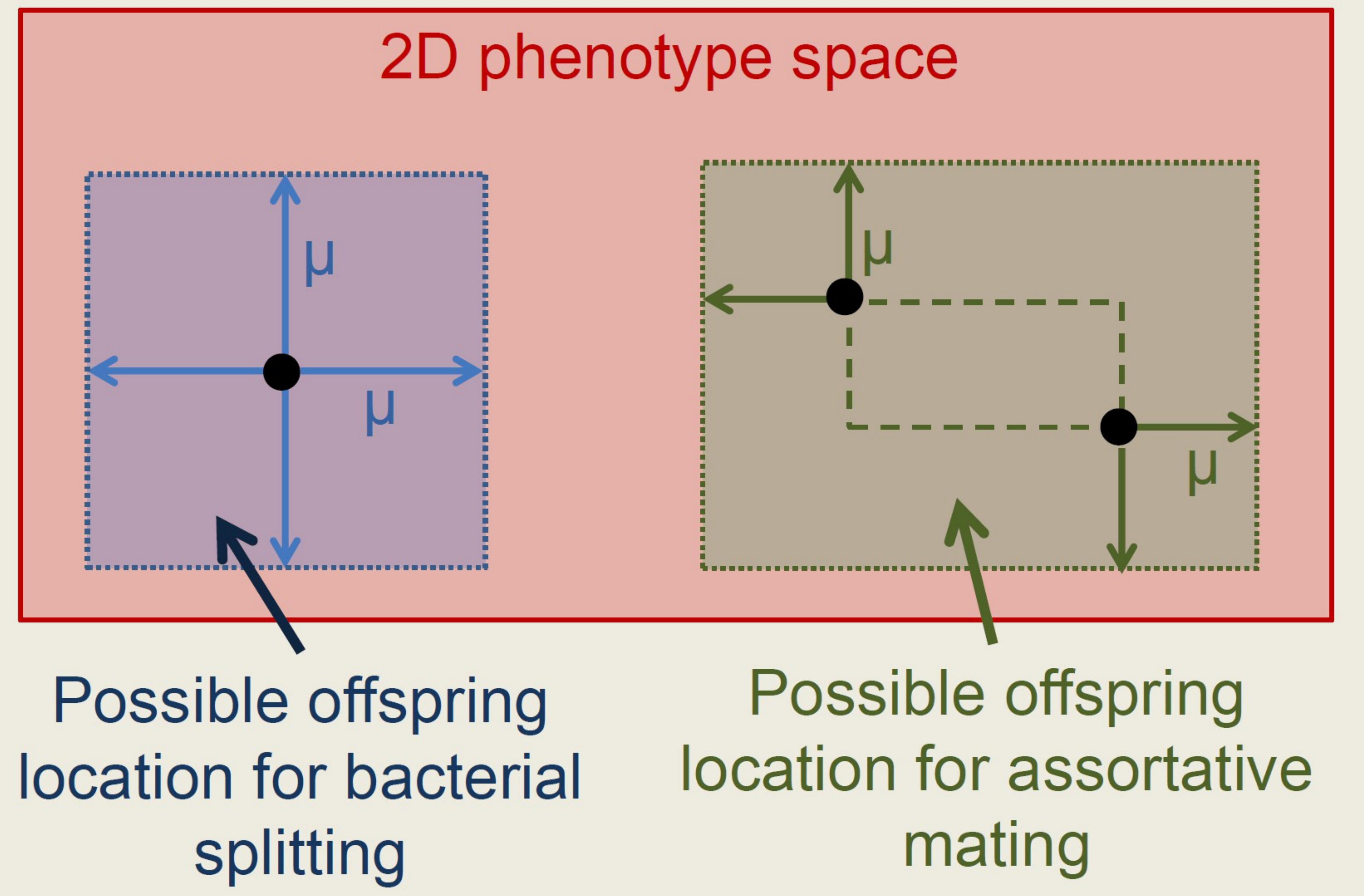}}
\caption{Sketch of the two reproduction schemes considered in our model.
The graphs show the positions of the offspring relative to their parents.
$\mu$ is the mutability in the evolutionary interpretation of the model.}
\label{fig:reproduction}
\end{figure}

The time evolution of the population from generation to generation consists
in three steps, (i) reproduction, (ii) competition death, and (iii) random death.
In the reproduction step, each organism produces $N_{fit}$ offspring. We
consider two different reproduction schemes, as sketched in Fig.\
\ref{fig:reproduction}.
In the asexual fission (or ``bacterial splitting'') scheme, each organism produces offspring
independently of the other organisms. The offspring coordinates are chosen
at random from a square of side $2\mu$ centered at the parent position.
In the evolutionary context, $\mu$ represents the
mutability. In the assortative mating scheme, each organism mates with its nearest neighbor.
More specifically, if organisms $a$ and $b$ are the nearest neighbors of each other, the pair
produce  $2\, N_{fit}$ offspring. If, on the other hand, organism $b$ is the nearest neighbor of $a$, but $c$ is the nearest neighbor of $b$, organism $b$ will produce  $N_{fit}$ offspring with $a$ and
another  $N_{fit}$ offspring with $c$. The offspring coordinates are randomly chosen from a rectangle around both parents, extended by $\mu$ in each direction.

After the reproduction step, the parent organisms are removed (die). The offspring
then undergo two consecutive death processes. First, if the distance between two
offspring organisms
is below the competition radius $\kappa$, one of them is removed at random.
This process simulates, e.g., the competition for limited resources.
 Second,
each surviving offspring dies with death probability $p$. These random deaths
are statistically independent of each other and model predation, accidents,
diseases, etc.
Offspring that survive both death processes form the parent population for the
next generation.
Note that the maximum number of organisms that can survive without competition-induced deaths,
$N_{max}$, corresponds to a hexagonal close packing of circles of radius $\kappa/2$.
Its value reads $N_{max}=(2/\sqrt{3})(L/\kappa)^2$.

If the parameters of this model, i.e., the number $N_{fit}$ of offspring per parent
as well as mutability $\mu$, competition radius $\kappa$, and death probability
$p$ do not depend on the position in space, no organism is given a fitness advantage.
This corresponds to neutral selection in the evolutionary context. In the present
paper, we only consider this case, but we will briefly discuss the effects
of a nontrivial
fitness landscape in the concluding section.

While we assume the model parameters to be uniform in space, we do investigate
variations in time (temporal disorder). They model temporal fluctuations in the
population's environment, e.g., climate fluctuations.
These fluctuations are global because they affect
all organisms in a given generation in the same way.
Specifically, we contrast the case of a constant, time-independent death
probability $p$ and the case in which $p$ varies randomly from
generation to generation.

\subsection{Monte Carlo simulations}

Close to the extinction transition, the population is expected to
fluctuate strongly on large length and time scales. This implies that large
systems and long simulation times are necessary to obtain accurate
results. We have therefore simulated large systems with sizes of up to
$1500\times 1500$. For a competition radius of $\kappa=0.25$,
this allows populations of up to 42 million organisms to exist without
competition-caused death.
 These systems are at least three orders of
magnitude larger than those used in earlier simulations
\cite{DeesBahar10,SKMB13,SKOB17}.

To tune the population between survival and extinction, we vary
the mutability $\mu$ while all other parameters are held fixed,
including the competition radius $\kappa=0.25$ and the number of offspring per
parent,  $N_{fit}=2$.
In simulations with time-independent death probability, we fix its
value at $p=0.3$. Temporal disorder is introduced by making the value
of $p$ for each generation an independent random variable drawn from
a binary distribution
\begin{equation}
W(p)=c\delta(p-p_h) +(1-c)\delta(p-p_l)
\label{eq:binary}
\end{equation}
with $p_h=0.5$, $p_l=0.1$, and $c=0.5$.

We perform two different types of simulations. The first kind starts from large populations
of about 80\% of the competition-limited maximum $N_{max}$ defined above (33 million organisms for the largest systems). These organisms are initially placed at random positions, and we follow their time evolution for up to 100,000 generations.
To obtain good statistics, we average over up to 40,000 individual runs for each parameter
set. The second kind
of simulation starts from a single organism and observes the growth of the population and
their spreading over space. Here, we average over more than $10^6$ individual runs.

\section{Extinction transition in time-independent environments}\label{sec:nodisorder}
\subsection{Asexual fission (bacterial splitting)}

\begin{figure}
\includegraphics[width=8.1cm]{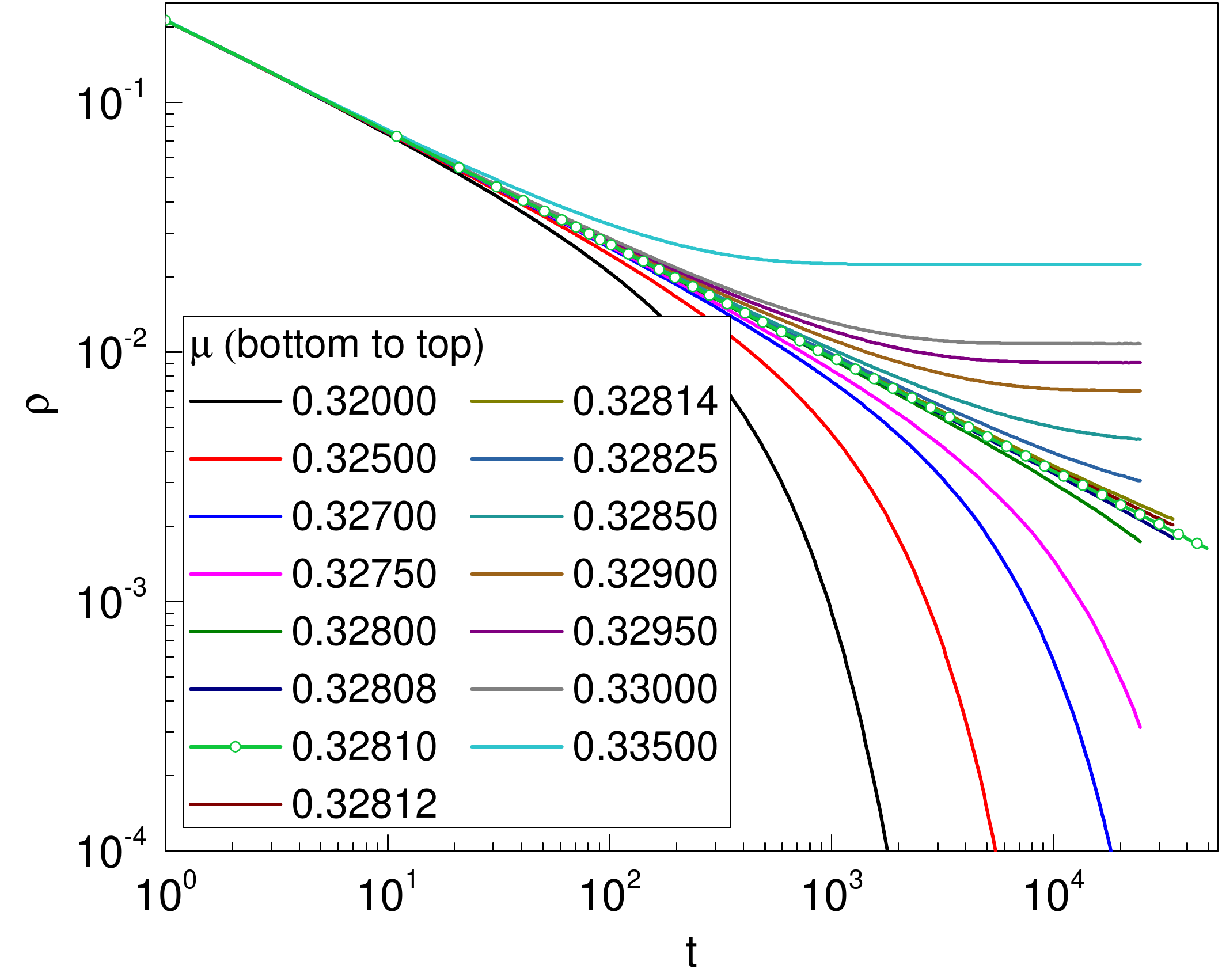}
\caption{Population density $\rho=N/N_{max}$ vs.\ time $t$ (in generations)
   for several $\mu$ close to the extinction
   transition for the bacterial splitting reproduction scheme. The data are averaged over 320 runs of a system having $L=1500$,
   $\kappa=0.25$, $N_{fit}=2$, and $p=0.3$. In each run, the initial population consists of
   $N_0=0.8\, N_{max} \approx 3.3\times 10^7$ organisms placed at random positions. The statistical error of $\rho$ does not exceed 10$^{-5}$. The critical curve, $\mu=\mu_c=0.32810$, is marked by circles.}
\label{fig:rhovst_clean}
\end{figure}

In order to find the extinction transition, we analyze the time dependence of the number $N$ of
organisms, starting from
large populations (80\% of $N_{max}$) for many different values of the mutability $\mu$,
as illustrated in Fig.\ \ref{fig:rhovst_clean}.
This figure shows a double-logarithmic plot of population density
$\rho=N/N_{max}$ vs.\ time $t$ (measured in generations) for parameters
$L=1500$, $\kappa=0.25$, $N_{fit}=2$, and $p=0.3$,

Three different regimes are clearly visible. For $\mu$ larger than some critical value
$\mu_c$, the population density initially decays but then settles onto a constant,
time-independent value. This is the active, surviving phase of the model.
For $\mu <\mu_c$, the population density decays to zero faster than a power law;
replotting the data in semilogarithmic form confirms an exponential decay.
This is the inactive phase in which even large populations go extinct quickly.
The two phases are separated by the critical point $\mu=\mu_c$ of the extinction transition
for which the population density decays like a power law.

To determine the value of $\mu_c$, we fit the population density to the power-law form
$\rho(t) \sim t^{-\delta}$. For $\mu=0.32810$, we obtain a high-quality fit (reduced $\chi^2 \approx 1.0$) over the time interval from $t=200$ to the longest times. The data for $\mu=0.32809$ and 0.32811 yield significantly worse fits. We thus conclude $\mu_c=0.32810(1)$ where the number in brackets indicates the error of the last digit. We have compared the results of different landscape sizes to ensure that this value is not affected by finite-size effects.
The power-law fit of the critical curve ($\mu=0.32810$) yields an exponent $\delta=0.4537$
with a very small statistical error of about $10^{-4}$. A larger contribution to the error stems from the remaining uncertainty in $\mu_c$. By comparing the fits for $\mu=0.32809$ and $\mu=0.32811$, we estimate
this error to be 0.002. Our final result for the decay exponent therefore
reads $\delta=0.454(2)$.

\begin{figure}
\includegraphics[width=8.1cm]{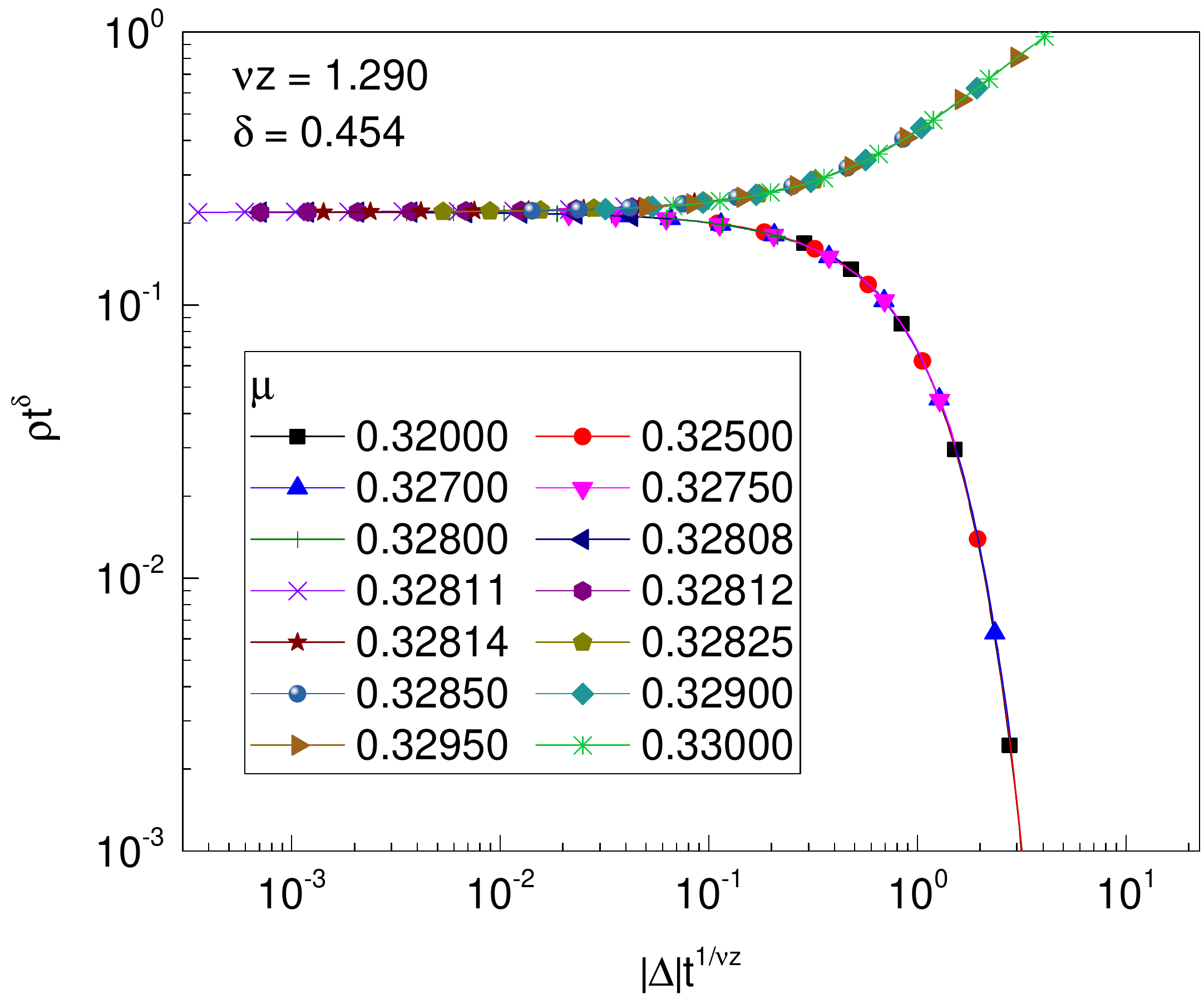}
\caption{Scaling plot of the density of Fig.\ \ref{fig:rhovst_clean} showing
$\rho t^\delta$ vs.\ $\Delta t^{1/(z \nu_\perp)}$ with $\delta=0.454$ and $z\nu_\perp =1.29$.
For clarity, only select data points are marked by symbols.}
\label{fig:scaling_clean}
\end{figure}

To analyze the off-critical behavior of the population density
(values of $\mu$ close to but different from $\mu_c$), we employ a scaling ansatz
appropriate for a nonequilibrium continuous phase transition (see, e.g., Ref.\ \cite{Hinrichsen00})
\begin{equation}
\rho(\Delta,t)=b^{-\beta/\nu_\perp}\rho(\Delta b^{1/\nu_\perp},tb^{-z})~.
\label{eq:rho_scaling_clean}
\end{equation}
Here, $\Delta=\mu-\mu_c$ denotes the distance from the transition point, and $b$ is an arbitrary
length scale factor. $\beta$, $\nu _\perp$, and $z$ are the order parameter, correlation length, and dynamical critical exponents, respectively. Setting the arbitrary scale factor to
$b=t^{1/z}$ yields the scaling form
\begin{equation}
\rho(\Delta,t)=t^{-\delta} X_\rho(\Delta t^{1/(z \nu_\perp)})
\label{eq:rho_scaling_clean_X}
\end{equation}
of the density, with $\delta=\beta/(z \nu_\perp)$ and scaling function $X_\rho$.
(At criticality, $\Delta=0$, eq.\ (\ref{eq:rho_scaling_clean_X}) turns into the
power-law decay $\rho \sim t^{-\delta}$ discussed earlier.)
The scaling form (\ref{eq:rho_scaling_clean_X}) can be used to find the exponent
combination $z\nu_\perp$: If one plots $\rho\, t^\delta$ vs.\ $\Delta t^{1/(z \nu_\perp)}$,
the data for all values of $\mu$ and $t$ should collapse onto a single master curve.
To perform this analysis, we fix the decay exponent at the value $\delta=0.454$ found earlier
and vary $z\nu_\perp$ until the best data collapse is obtained. This yields $z\nu_\perp=1.29$. The collapse for $z\nu_\perp=1.27$ or 1.31 is of visibly lower quality. We therefore arrive at
the final estimate $z\nu_\perp=1.29(2)$. The resulting scaling plot
is shown in Fig.\ \ref{fig:scaling_clean}.

\begin{figure}
\includegraphics[width=8.1cm]{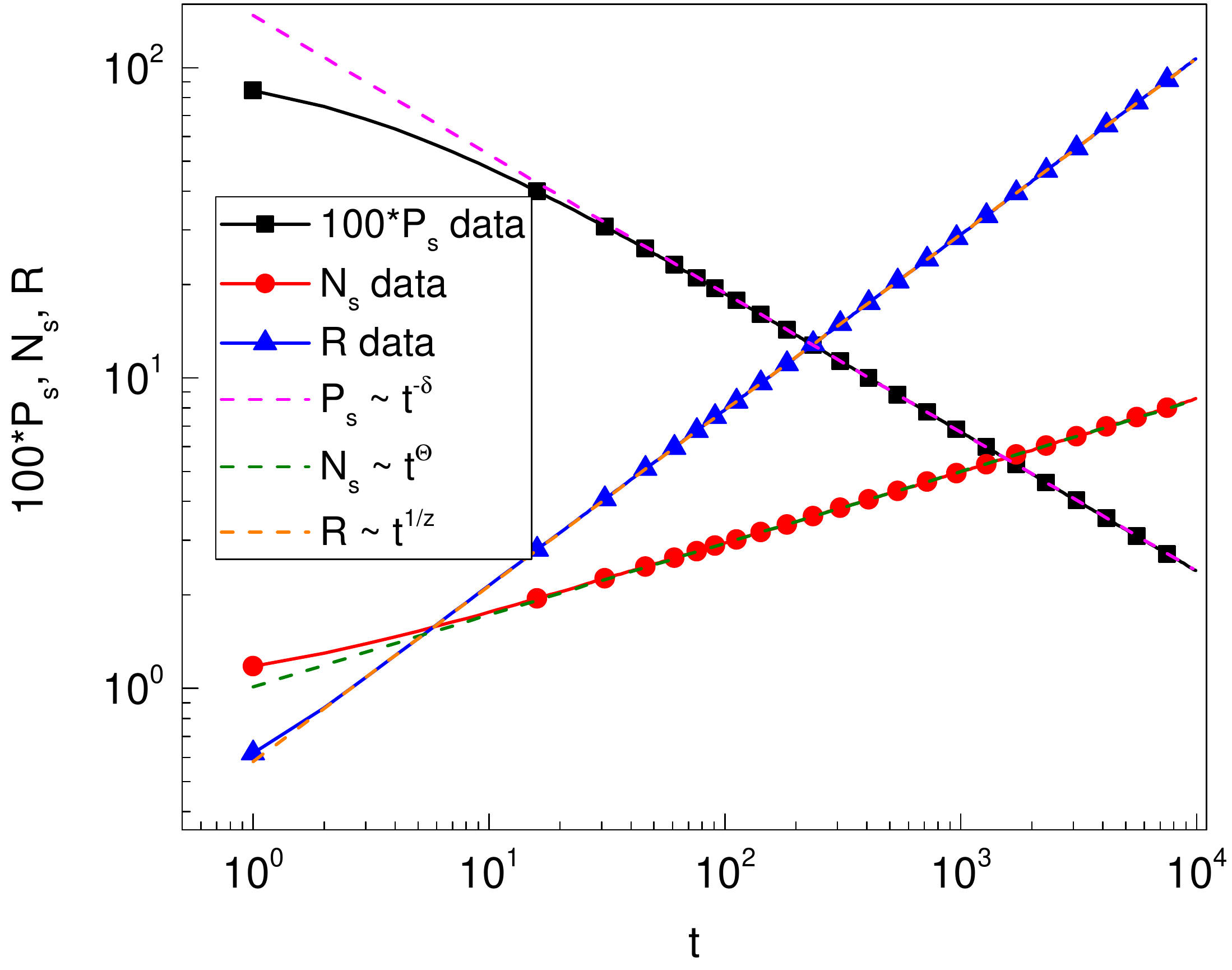}
\caption{Survival probability $P_s$, number of organisms $N_s$, and radius $R$
vs.\ time $t$ for $\mu=\mu_c=0.32809$. The data are averages over $2\times 10^6$
 independent runs, each starting from a single organism. The resulting statistical errors are much smaller than the symbol size. The other parameters,
 $L=1500$, $\kappa=0.25$, and $N_{fit}=2$), are identical to those used in Fig.\ \ref{fig:rhovst_clean}.
 The dashed lines represent power-law fits.}
\label{fig:spreading_clean}
\end{figure}

In addition to the simulations that follow the time evolution of large populations, we
also perform runs that start from a single organism in the center of the landscape
and observe the spreading of the
population through space. Specifically, we analyze the survival probability $P_s(t)$,
i.e., the probability that the population has not gone extinct at time $t$. We also measure
the average number $N_s(t)$ of organism in the population at time $t$ as well as the
mean-square radius $R(t)$ of the cloud of organisms in space. Right at the extinction
transition, these quantities are expected to follow the power laws
\begin{equation}
P_s(t) \sim t^{-\delta}, \quad N_s(t) \sim t^\Theta, \quad R(t) \sim t^{1/z}~.
\label{eq:spreading_clean}
\end{equation}
From the quality of the fits of our data to these power laws, we determine the
critical point to be $\mu_c=0.32809(2)$, in good agreement with the estimate
obtained from the density $\rho(t)$ above. The resulting exponent values read
$\delta=0.449(7)$, $\Theta=0.235(10)$, and $z=1.76(1)$. The value of $\delta$
is slightly less precise but agrees with the value obtained from $\rho(t)$.
Figure \ref{fig:spreading_clean} shows $P_s(t)$, $N_s(t)$, and $R(t)$ at the
critical point.

The values of the critical exponents resulting from our simulations are summarized in
Table \ref{table:exponents_clean}.
\begin{table}
\renewcommand*{\arraystretch}{1.2}
\begin{tabular}{cccc}
\hline\hline
Exponent     & this work  & DP, Ref.\ \cite{Dickman99} & DP, Ref.\ \cite{VojtaFarquharMast09}\\
\hline
$\delta$     & 0.454(2)  & 0.4523(10) & 0.4526(7)\\
$\Theta$     & 0.235(10) & 0.2293(4)   & 0.233(6)\\
$z$          & 1.76(1)   & 1.767(1)    & 1.757(8)\\
$z\nu_\perp$ & 1.29(2)   & 1.292(4)    & 1.290(4)\\
\hline
$\nu_\perp$  & 0.73(2)   &             & \\
$\beta$      & 0.586(12) &             & \\
\hline\hline
\end{tabular}
\caption{Critical exponents for the bacterial splitting model in a
\emph{time-independent} environment compared to
high-accuracy results for the two-dimensional directed percolation (DP) universality class.
The exponents above the horizontal line are directly measured in our simulations, the ones
below are computed using exponent relations.}
\label{table:exponents_clean}
\end{table}
They fulfill the hyperscaling relation \cite{GrassbergerdelaTorre79} $\Theta+2\delta =d/z$ within their error bars.
For comparison, Table  \ref{table:exponents_clean} also shows the exponents obtained in two
high-accuracy studies of the contact process \cite{Dickman99,VojtaFarquharMast09}
which is a paradigmatic model in the directed percolation \cite{GrassbergerdelaTorre79}
universality class.
Our exponents agree very well with the directed percolation values, providing
strong evidence for the extinction transition in our model to belong to this universality class.
This is in agreement with a conjecture by Janssen and Grassberger \cite{Janssen81,Grassberger82},
according to which all absorbing state transitions with a scalar order parameter, short-range interactions, and no extra symmetries or conservation laws belong to the directed percolation universality class.

\subsection{Assortative mating}

In addition to the bacterial splitting case, we also perform simulations using the
assortative mating reproduction scheme for which the offspring is located in a rectangle
around the organism and its nearest neighbor, expanded by the mutability $\mu$
(see Fig.\ \ref{fig:reproduction}). It is important to point out that assortative mating
introduces a kind of long-range interaction between the organisms because the distance between an organism and
its nearest neighbor can become arbitrarily large. It is thus not clear whether or not one should
expect the extinction transition to belong to the directed percolation universality class.

\begin{figure}
\includegraphics[width=8.1cm]{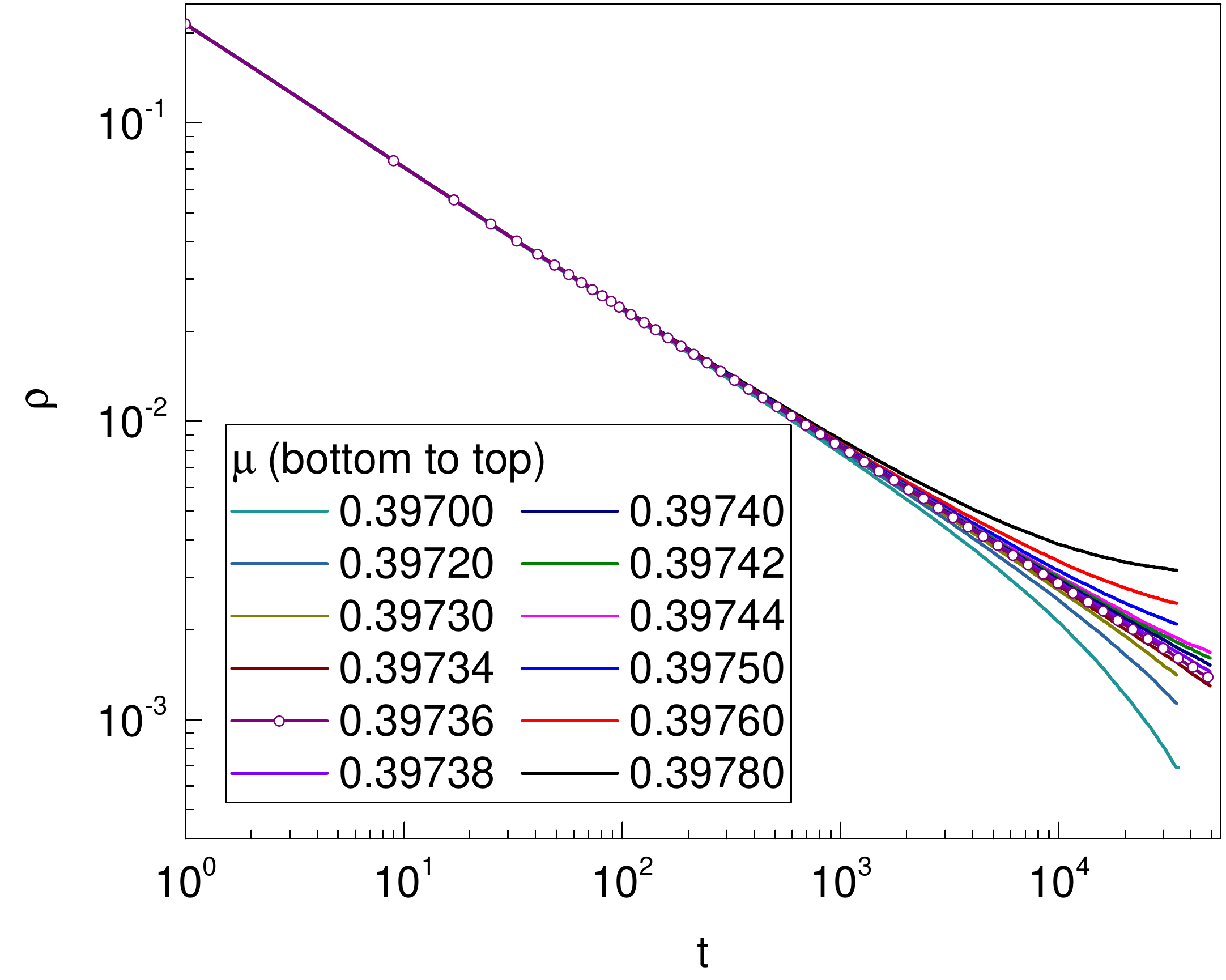}
\caption{Population density $\rho=N/N_{max}$ vs.\ time $t$ (in generations)
   for several $\mu$ close to the extinction
   transition for the assortative mating reproduction scheme. The data are averages over 200 runs with parameters $L=1500$,
   $\kappa=0.25$, $N_{fit}=2$, and $p=0.3$. In each run, the initial population consists of
   $N_0=0.8\, N_{max} \approx 3.3\times 10^7$ organisms placed at random positions. The statistical error of $\rho$ does not exceed 10$^{-5}$. The critical curve, $\mu=\mu_c=0.39736$, is marked by circles.}
\label{fig:assortative_clean_rhovst}
\end{figure}

Figure \ref{fig:assortative_clean_rhovst} shows the time evolution of the population
density $\rho$ starting from a large population (80\% of $N_{max}$) for many different
values of $\mu$. The parameters $L=1500$, $\kappa=0.25$, $N_{fit}=2$, and $p=0.3$  are chosen to be identical to the bacterial splitting simulations discussed in the last section.
To determine the critical point, we fit the population density to the power law $\rho(t) \sim
t^{-\delta}$. The best power law is found for $\mu=0.39736$ (high-quality fit with reduced $\chi^2
\approx 1$ from $t=500$ to the longest times) while the fits for $\mu=0.39734$ and $0.39738$ are of lower quality. We therefore conclude that the critical point is at $\mu_c=0.39736(2)$.
The exponent resulting from the power-law fit of the critical curve is $\delta=0.459(5)$. The
majority of the error stems from the uncertainty in $\mu_c$; the statistical error is only about $2\times 10^{-4}$.

\begin{figure}
\includegraphics[width=8.1cm]{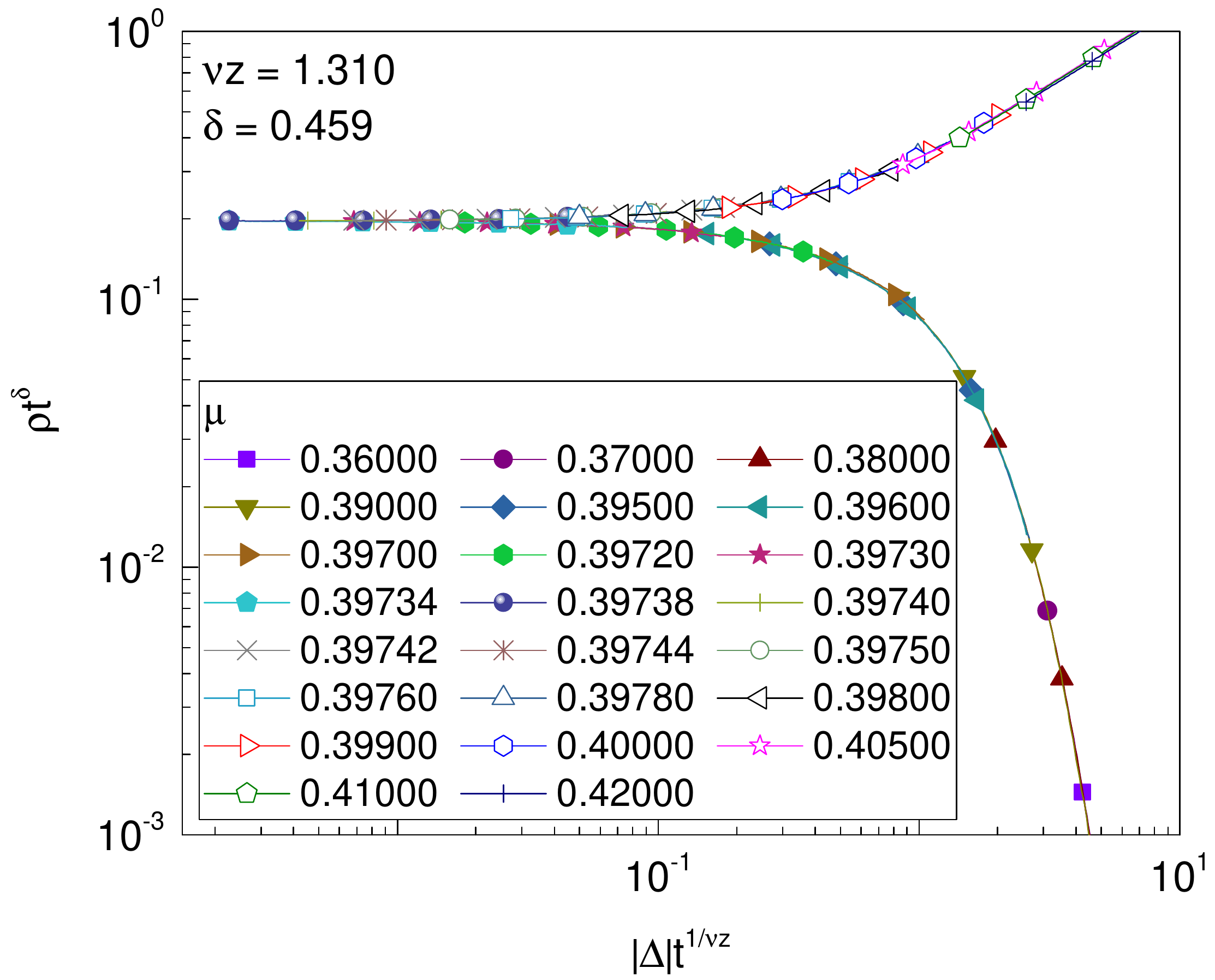}
\caption{Scaling plot of the population density for the assortative mating case showing
$\rho t^\delta$ vs.\ $\Delta t^{1/(z \nu_\perp)}$ with $\delta=0.459$ and $z\nu_\perp =1.31$.
For clarity, only select data points are marked by symbols.}
\label{fig:assortative_clean_scaling}
\end{figure}

To study the off-critical behavior, we again perform a scaling analysis of
the population density according to eq.\ (\ref{eq:rho_scaling_clean_X}). We fix $\delta$ at the value found above, $\delta=0.459$, and vary $z\nu_\perp$ until we find the best data collapse. This yields
$z\nu_\perp=1.31(2)$. The resulting high-quality scaling plot is shown in Fig.\
\ref{fig:assortative_clean_scaling}.

The critical exponents of the extinction transition for the assortative mating reproduction scheme,
$\delta=0.459(5)$ and $z\nu_\perp=1.31(2)$ are very close to the directed percolation
exponents listed in Table \ref{table:exponents_clean}. Specifically, $z\nu_\perp$ agrees
within the error bars with the directed percolation value while $\delta$ is very slightly
too high (the errors bars almost touch, though). We believe the small deviation stems from finite-size
effects which are significantly stronger than in the bacterial splitting case.
Simulations for a smaller landscape of $L=500$ yield an exponent value of $\delta=0.466$
which suggests that the true (infinite system) exponent will be slightly
below 0.459(5).

We conclude that the extinction transition in the assortative mating case also
belongs to the directed percolation universality class despite the potential long-range
interactions introduced by the assortative mating. This may be caused by the fact that
almost all mating pairs have short distances. A long mating distance occurs only if a
\emph{single} organism is far way from all others.\footnote{In contrast,
for anomalous directed percolation with long-range spreading, every organism can produce offspring
far away from its own position \cite{JOWH99,HinrichsenHoward99}.}
In this case, the positions of its
offspring (and offspring of offspring) move quickly towards those of the other organisms. The net
effect is not that different from the isolated organism simply dying.

Alternatively, the directed percolation exponents could hold in a transient time regime only
while the long-range interactions cause the true asymptotic exponents to differ from directed percolation.
However, our simulations of large populations for fairly long times do not show any indications
of a crossover away from directed percolation behavior. Determining rigorously whether or not the
assortative mating model belongs to the directed percolation universality class
therefore remains a task for the future.

\section{Extinction transition fluctuating environments}\label{sec:disorder}

We now turn to the effects of temporal environmental fluctuations, i.e., temporal disorder,
on the extinction transition. To this end, we perform simulations (using the
 bacterial splitting reproduction scheme) in which the death probability
$p$ varies randomly from generation to generation. Specifically, the value of $p$ for each
generation is an independent random variable drawn from the binary distribution
(\ref{eq:binary}) with $p_h=0.5$, $p_l=0.1$, and $c=0.5$.

\begin{figure}
\includegraphics[width=8.1cm]{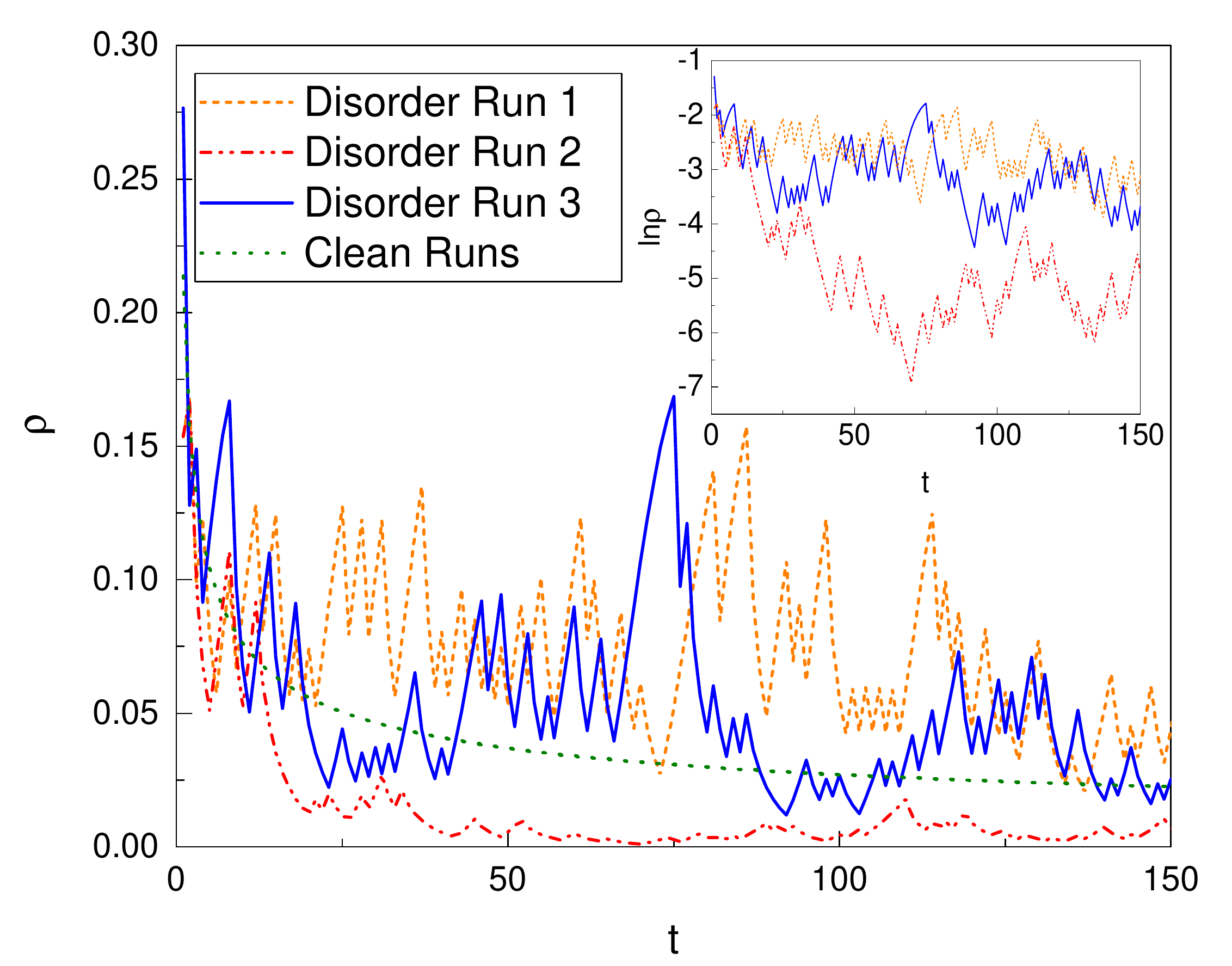}
\caption{Time evolution of the population density for three individual runs,
each with a different temporal disorder configuration. The runs start from a
population of $N_0=0.8\, N_{max}\approx 1.5\times 10^7$ organisms and use parameters
$L=1000$, $\mu=0.4129$, $\kappa=0.25$, $N_{fit}=2$, as well as $p=0.1$ or 0.5 with 50\% probability.
Also shown are three runs for the analogous system without temporal disorder, having $p=0.3$
and $\mu=0.32810$.
The inset shows the disordered runs on a logarithmic scale.}
\label{fig:individual}
\end{figure}

The environmental noise induces enormous population fluctuations even for large populations.
This is illustrated in Fig.\ \ref{fig:individual} which shows
the time evolutions of three individual populations subject to different realizations
of the temporal disorder.
All three populations initially have the same large population of $N_0=0.8\, N_{max}\approx 1.5\times 10^7$ organisms. Nonetheless, after fewer than 100 generations, the populations already differ
by more than two orders of magnitude. For comparison, the figure also shows three populations \emph{not} subject to environmental noise but only to the demographic (Monte Carlo) noise of the stochastic time evolution. In this case, the populations curves are smooth and differ from each other by only a few percent. They are thus indistinguishable in  Fig.\ \ref{fig:individual}.
Environmental noise is much more efficient in inducing population fluctuations than intrinsic demographic noise because it affects the entire population in the same way while the demographic noise
acting on one organism is independent from that acting on other organisms.

\begin{figure}
\includegraphics[width=8.1cm]{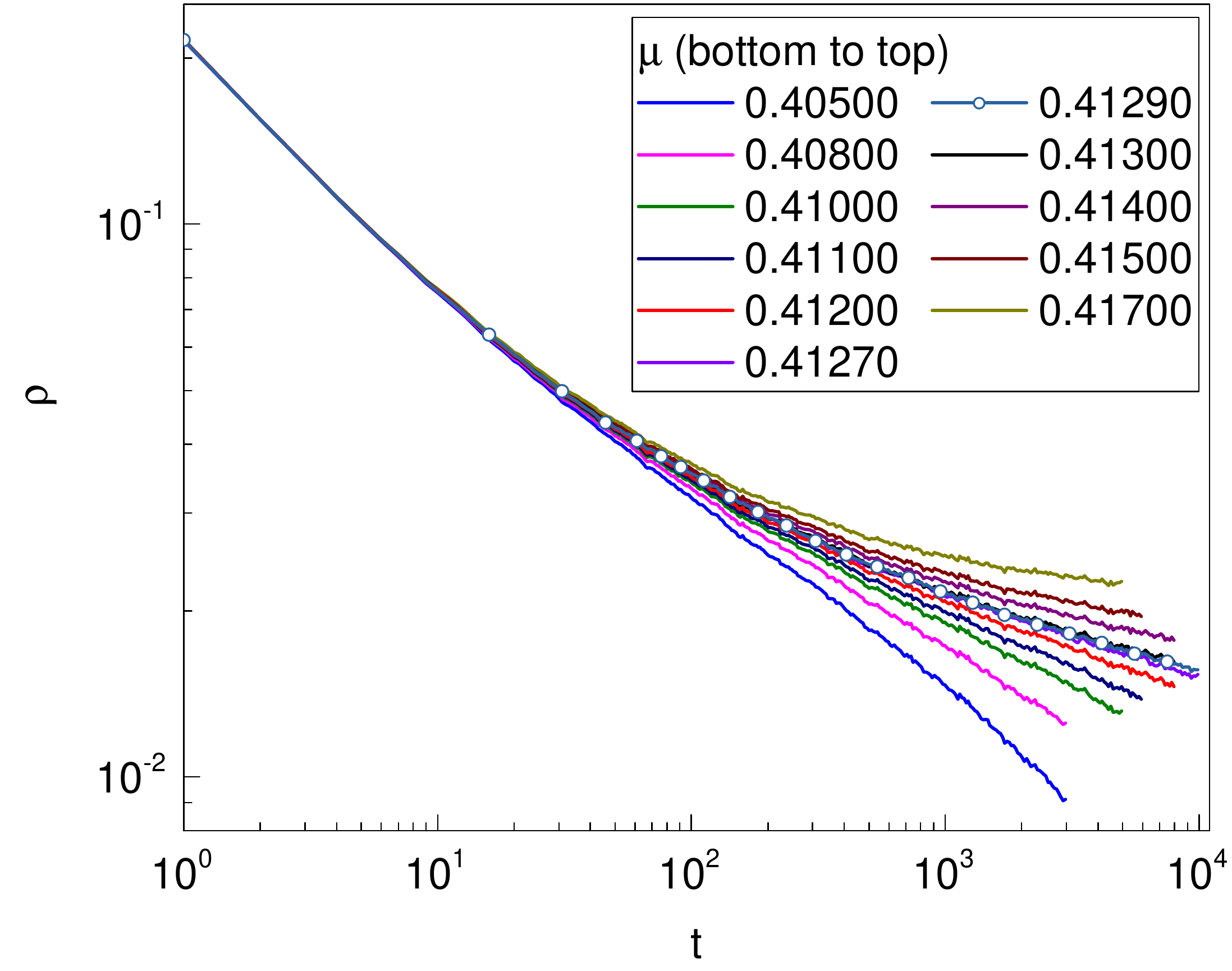}
\caption{Double logarithmic plot of the average population density $\rho$ vs.\ time $t$ (in generations)
   in the presence of temporal disorder. The data are averages over 20,000 to 40,000 runs, each having a different disorder realizations and parameters $L=1000$,
   $\kappa=0.25$, $N_{fit}=2$, and $p=0.1$ or 0.5 with 50\% probability. The initial population consists of
   $N_0=0.8\, N_{max} \approx 1.5\times 10^7$ organisms placed at random positions. The statistical error of $\rho$ does not exceed $2\times 10^{-4}$. }
\label{fig:rhovst_disordered_loglog}
\end{figure}

To investigate the extinction transition quantitatively, we now analyze the time evolution of
the average population density. Figure  \ref{fig:rhovst_disordered_loglog} presents
a double logarithmic plot of $\rho$ vs.\ $t$, averaged over 20,000 to 40,000
temporal disorder realizations.
As in the clean case, Fig.\ \ref{fig:rhovst_clean}, we can identify the active, surviving phase in which the average population density $\rho$ approaches a nonzero constant for long times as well as the inactive phase, in which the density decays faster than a power law with time. However, the curve separating the two phases is clearly not a power law. In fact, none of the curves can be described by a power law over any appreciable time interval. This suggests that the extinction transition in the  presence of temporal disorder, i.e., environmental noise,  is unconventional.

\begin{figure}
\includegraphics[width=8.1cm]{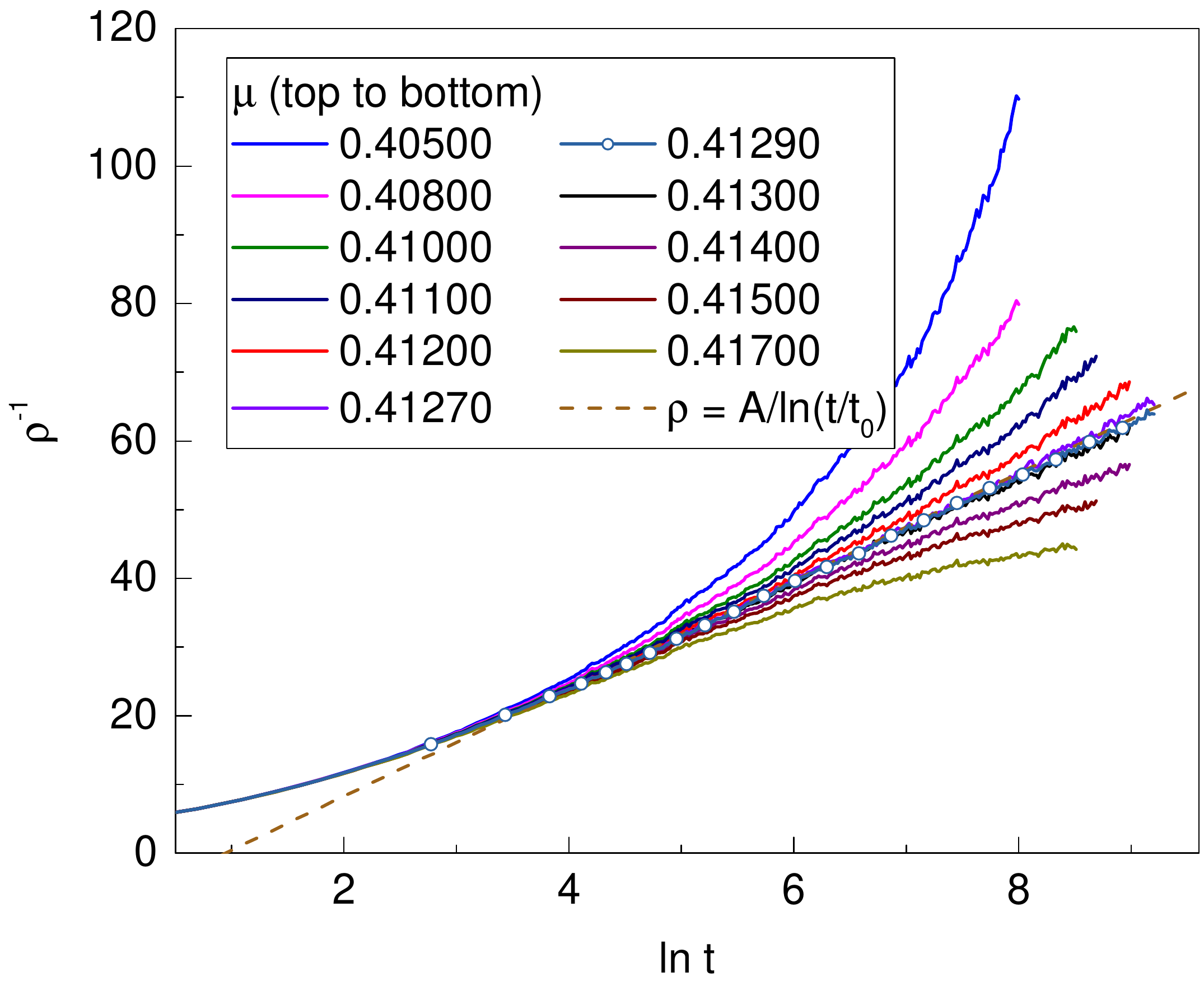}
\caption{Inverse average population density $\rho^{-1}$ vs.\ $\ln(t)$. The
         parameters are identical to Fig.\ \ref{fig:rhovst_disordered_loglog}.
         The dashed line is a fit of the critical curve, $\mu=0.4129$, to the
         logarithmic decay law  (\ref{eq:rho_disorder}).  }
\label{fig:rhovst_disordered}
\end{figure}

Recently, Vojta and Hoyos \cite{VojtaHoyos15} developed a real-time (``strong-noise'') renormalization group
approach to absorbing state transitions with temporal disorder. This theory predicts
an exotic ``infinite-noise'' critical point. It is characterized by a slow, logarithmic
decay of the average population density at the transition point,
\begin{equation}
\rho(t) \sim [\ln(t/t_0)]^{-\bar\delta}
\label{eq:rho_disorder}
\end{equation}
where $\bar\delta=1$, and $t_0$ is a non-universal time scale.
To test this prediction, we plot in Fig.\ \ref{fig:rhovst_disordered}
the density data in the form $\rho^{-1}$ vs.\ $\ln(t)$. In this plot,
the predicted critical behavior (\ref{eq:rho_disorder}) yields
a straight line, independent of the value of $t_0$.
The figure shows that the data for $\mu=0.4129$ indeed follow a straight line for more
than two orders of magnitude in time. Accordingly, a fit of these data
to eq.\ (\ref{eq:rho_disorder}) from $t=50$ to the longest times ($t=10,000$) is
of high quality (reduced $\chi^2 \approx 1$), significantly better than the fits
for $\mu=0.4127$ and $\mu=0.4130$. (The resulting $t_0$ value is $t_0=2.572$.)
Consequently, we conclude that the critical point is located at $\mu=0.4129(2)$
and fulfills the predicted infinite-noise behavior.

Based on the renormalization group approach \cite{VojtaHoyos15}, a heuristic scaling
theory was developed in Ref.\ \cite{BarghathiVojtaHoyos16}.
The scaling ansatz for the average population density reads
\begin{equation}
\rho(\Delta, t) = (\ln b)^{-\beta/\bar\nu_\perp} \rho[\Delta (\ln b)^{1/\bar\nu_\perp},tb^{-z}]~,
\label{eq:rho_scaling_disorder}
\end{equation}
and the predicted exponent values are $\beta=0.5$, $\bar\nu_\perp=0.5$, and $z=1$.
Setting the length scale factor $b=(t/t_0)^{1/z}$ leads to the scaling form
\begin{equation}
\rho(\Delta, t) = [\ln (t/t_0)]^{-\beta/\bar\nu_\perp} X_\rho\{\Delta [\ln (t/t_0)]^{1/\bar\nu_\perp}\}
\label{eq:rho_scaling_disorder_X}
\end{equation}
where the non-universal time scale $t_0$ is necessary because logarithms are not scale free.
Interestingly, due to the logarithmic time dependence, the value of the dynamical exponent $z$ does not appear in (\ref{eq:rho_scaling_disorder_X}).
For $\Delta=0$, we recover the logarithmic decay (\ref{eq:rho_disorder}) with $\bar\delta=\beta/\bar\nu_\perp=1$.

\begin{figure}
\includegraphics[width=8.1cm]{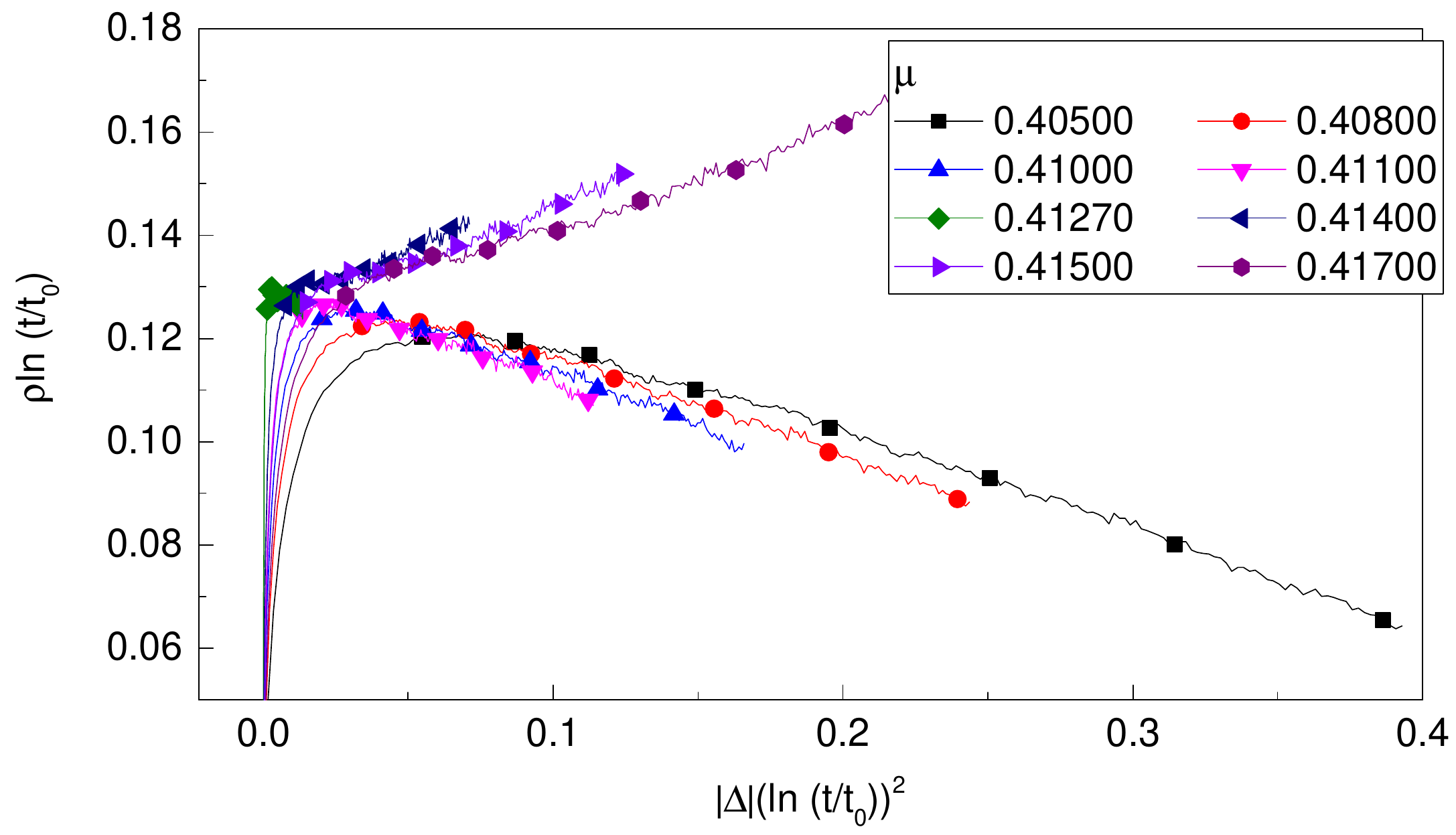}
\caption{Scaling plot of the density of Fig.\ \ref{fig:rhovst_disordered} showing
$\rho\, \ln (t/t_0) $ vs.\ $\Delta\,[\ln(t/t_0)]^2$ with $t_0=2.572$.
For clarity, only select data points are marked by symbols.}
\label{fig:disordered_scaling_nu=2_t1=t0}
\end{figure}

\begin{figure}
\includegraphics[width=8.1cm]{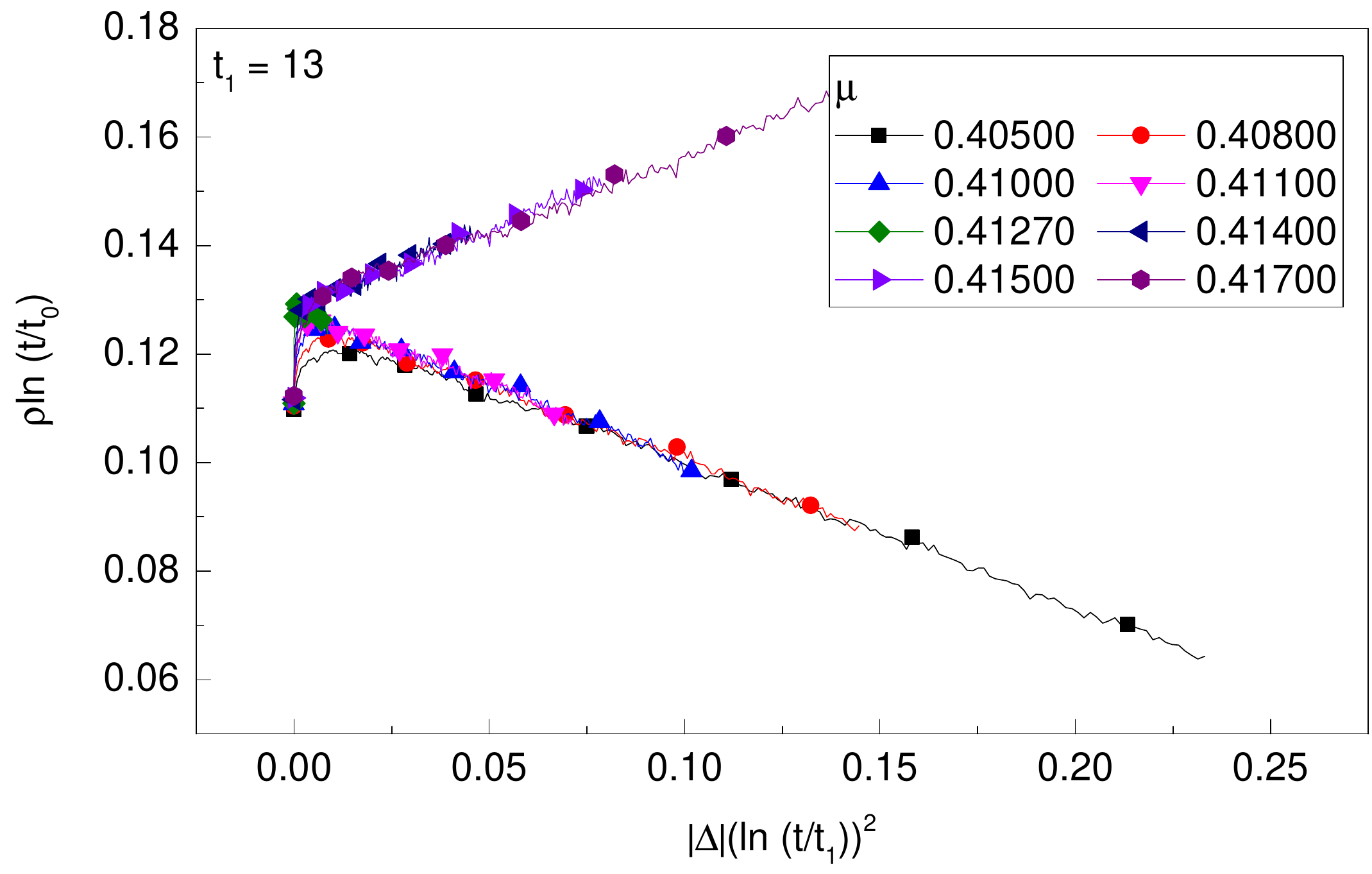}
\caption{Scaling plot of the density of Fig.\ \ref{fig:rhovst_disordered} showing
$\rho\, \ln (t/t_0) $ vs.\ $\Delta\,[\ln(t/t_1)]^2$ with $t_0=2.572$ and $t_1=13.0$.}
\label{fig:disordered_scaling_nu=2_t1net0}
\end{figure}

\begin{figure}
\includegraphics[width=8.1cm]{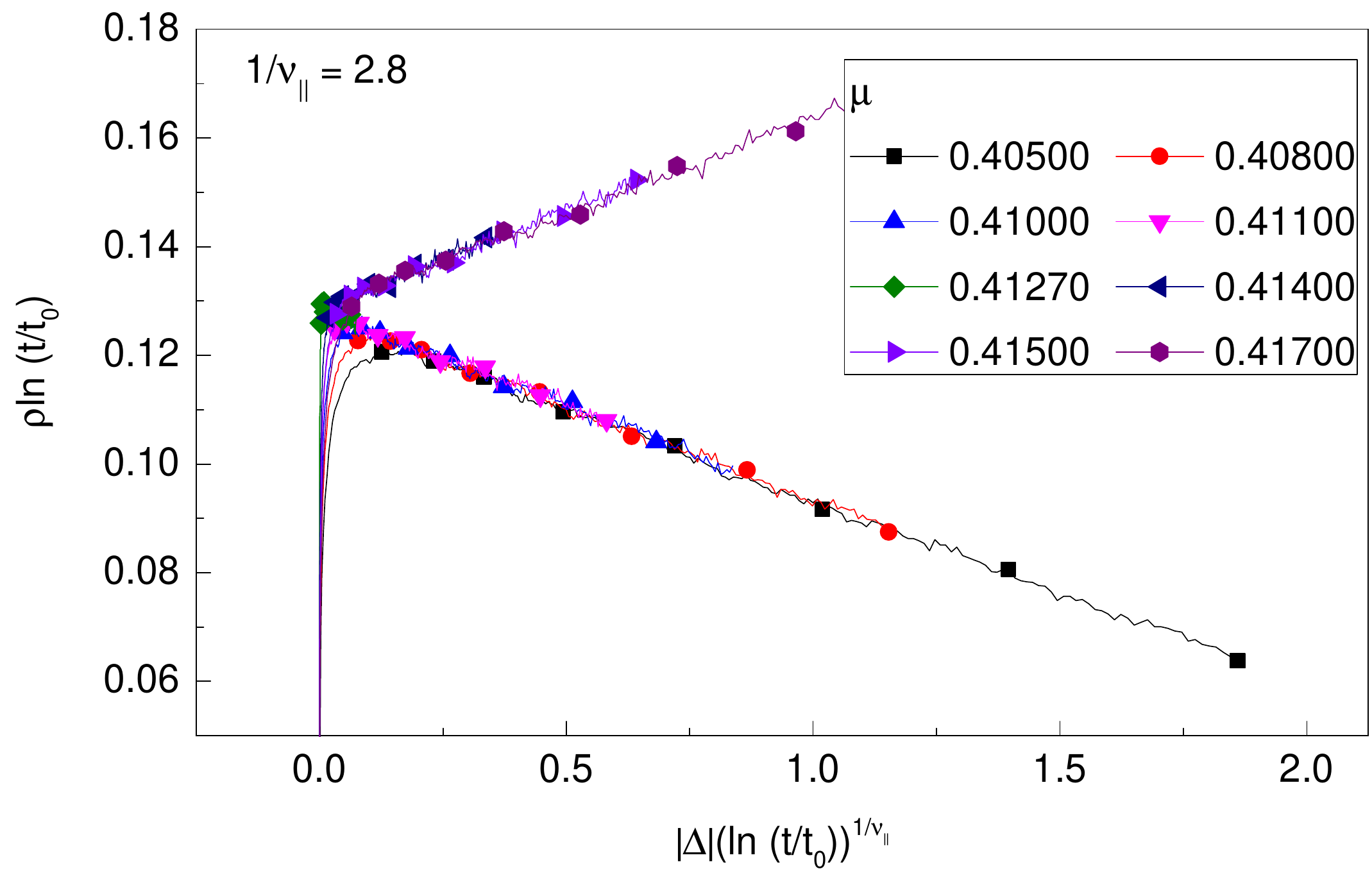}
\caption{Scaling plot of the density of Fig.\ \ref{fig:rhovst_disordered} showing
$\rho\, \ln (t/t_0) $ vs.\ $\Delta\,[\ln(t/t_0)]^{2.8}$ with $t_0=2.572$.}
\label{fig:disordered_scaling_nu=28_t1=t0}
\end{figure}

The scaling form implies that the data for all $\mu$ and $t$ should collapse onto
a master curve if one graphs $\rho\, \ln (t/t_0) $ vs.\ $\Delta\,[\ln(t/t_0)]^2$.
The corresponding scaling plot, using the value $t_0=2.572$ found earlier
is presented in Fig.\ \ref{fig:disordered_scaling_nu=2_t1=t0}.
Clearly, the data collapse is not particularly good. A better collapse
can be achieved if  we allow the microscopic time scale on the $x$-axis to
differ from $t_0$. Within the scaling description of the critical point
this corresponds to a subleading correction to the leading scaling behavior.
Figure \ref{fig:disordered_scaling_nu=2_t1net0} shows a scaling plot of
$\rho\, \ln (t/t_0) $ vs.\ $\Delta\,[\ln(t/t_1)]^2$ where $t_1$ is independent
of $t_0$.
The value $t_1=13.0$ leads to nearly perfect data collapse. Alternatively,
a good collapse can be obtained by varying the exponent in the scaling combination
$\Delta \ln (t/t_0)]^{1/\bar\nu_\perp}$ on the $x$-axis. Figure \ref{fig:disordered_scaling_nu=28_t1=t0}
demonstrates that $1/\bar\nu_\perp=2.8$ results in a good collapse.
Distinguishing between these two scenarios, viz., (i) critical exponents as predicted by
the renormalization group theory, but with corrections to scaling and (ii) an exponent value
that differs from the renormalization group theory, requires data over significantly longer
times that are presently beyond our numerical means.

In addition to the simulations studying the time evolution of large populations, we
also perform runs that start from a single organism and observe the spreading of the
population through space. The scaling theory of Ref.\ \cite{BarghathiVojtaHoyos16}.
leads to the following predictions for the time dependencies of the survival probability
$P_s$, the number of organisms $N_s$, and the radius $R$ of the organism cloud at criticality:
\begin{eqnarray}
P_s(t) &\sim& [\ln(t/t_0)]^{-1}
\label{eq:disordered_Ps}\\
N_s(t) &\sim& t^2 [\ln(t/t_0)]^{-y_N}
\label{eq:disordered_Ns}\\
R(t) &\sim& t [\ln(t/t_0)]^{-y_R}
\label{eq:disordered_R}
\end{eqnarray}
Interestingly, except for the subleading logarithmic corrections, these time dependencies are the same as in the active, surviving phase where the population spreads ballistically, $R \sim t$ and $N_s \sim t^2$.
The theory does not predict the values of the exponents $y_N$ and $y_R$ governing the logarithmic
corrections.

\begin{figure}
\includegraphics[width=8.1cm]{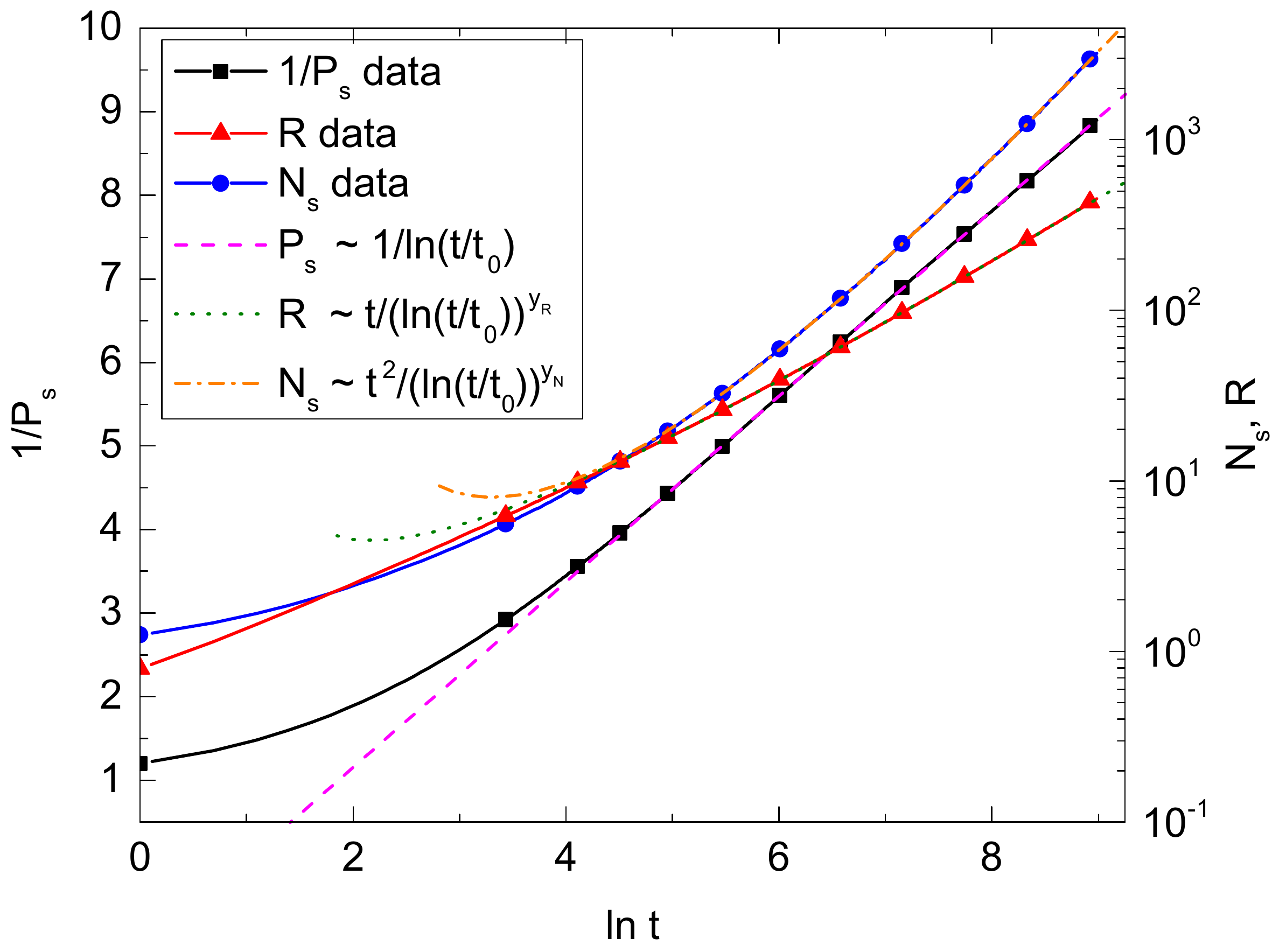}
\caption{Survival probability $P_s$, number of organisms $N_s$, and radius $R$
vs.\ time $t$ for $\mu=\mu_c=0.4129$ $\kappa=0.25$, $N_{fit}=2$, and
$p=0.1$ or 0.5 with 50\% probability. The data are averages over $150,000$
realizations of the temporal disorder with 10 independent runs per realization.
Each run starts from a single organism in the center of the landscape.
The statistical errors of all data are much smaller than the symbol size.
The dashed and dotted lines represent fits to the theoretical predictions
(\ref{eq:disordered_Ps}), (\ref{eq:disordered_Ns}), and (\ref{eq:disordered_R}).}
\label{fig:disordered_spreading}
\end{figure}

Figure \ref{fig:disordered_spreading} presents the results of our spreading simulations
at the critical point, $\mu=0.4129$. The data are averages over 150,000 realizations of the
temporal disorder with 10 attempts of growing the population per realization.
Each attempt starts from a single organism in the center of a landscape of size $L=1500$.
The other parameters are identical to those used earlier, $\kappa=0.25$, $N_{fit}=2$, and
$p=0.1$ or 0.5 with 50\% probability.
The figure shows that the data for times larger than about 100 are well described
by the theoretical predictions. Accordingly, fits to eqs.\
(\ref{eq:disordered_Ps}), (\ref{eq:disordered_Ns}), and (\ref{eq:disordered_R})
yield reduced $\chi^2 \approx 1$. The resulting values of the exponents
$y_N$ and $y_R$ have small statistical errors of about $3\times 10^{-2}$.
However, as they stem from subleading terms, they are sensitive towards small
changes of the fit interval. Taking this into account, we arrive at the
estimates $y_N=3.9(4)$ and $y_R=1.1(2)$. These exponents fulfill the equality
\cite{BarghathiVojtaHoyos16} $y_N=2\bar\delta +dy_R$ with $\bar\delta=1$.
However, they differ from the values found in simulations of the two-dimensional
contact process \cite{BarghathiVojtaHoyos16}. This can either mean that these subleading
exponents are non-universal or that our values have not yet reached the asymptotic
long-time regime. Distinguishing these possibilities will require simulations
over much longer time intervals.

Finally, we determine how the mean time to extinction (or average life time) of a
population depends on its size. In time-independent environments, a finite population that is
nominally on the active, surviving side of the extinction transition can decay only via a rare
fluctuation of the demographic (Monte-Carlo) noise. The probability of such a rare
event decreases exponentially with population size, leading to an exponential dependence of
the lifetime on the population size.

In the presence of (uncorrelated) temporal disorder, the extinction probability is enhanced
due to rare, unfavorable fluctuations of the environment. Within space-independent
(mean-field) models of population dynamics, this leads to a power-law relation (rather than an exponential one) between
life time and population size \cite{Leigh81,Lande93}. Vazquez et al. \cite{VBLM11} pointed
out that this power-law behavior can be understood as the temporal analog of the
Griffiths singularities known in spatially disordered systems. They dubbed the parameter region
where the power-law behavior occurs the temporal Griffiths phase.

\begin{figure}
\includegraphics[width=8.1cm]{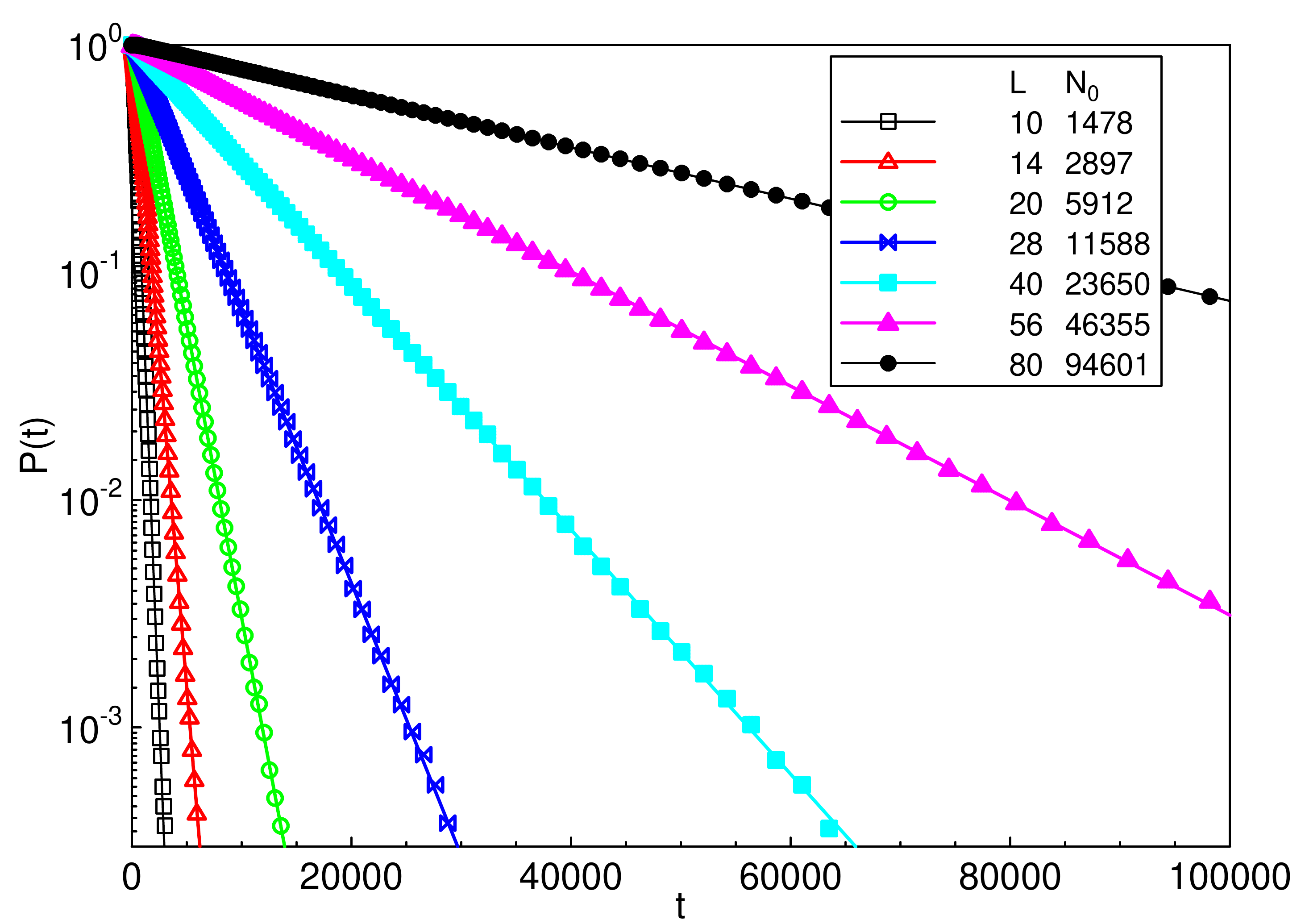}
\caption{Semi-logarithmic plot of the survival probability $P$, vs.\ time $t$ for several population sizes,
$\mu=0.420$, $\kappa=0.25$, $N_{fit}=2$, and $p=0.1$ or 0.5 with 50\% probability.
The data are averages over 100,000 runs, each with a different realization of the temporal
disorder. The lines are exponential fits.}
\label{fig:ptmu0420}
\end{figure}

\begin{figure}
\includegraphics[width=8.1cm]{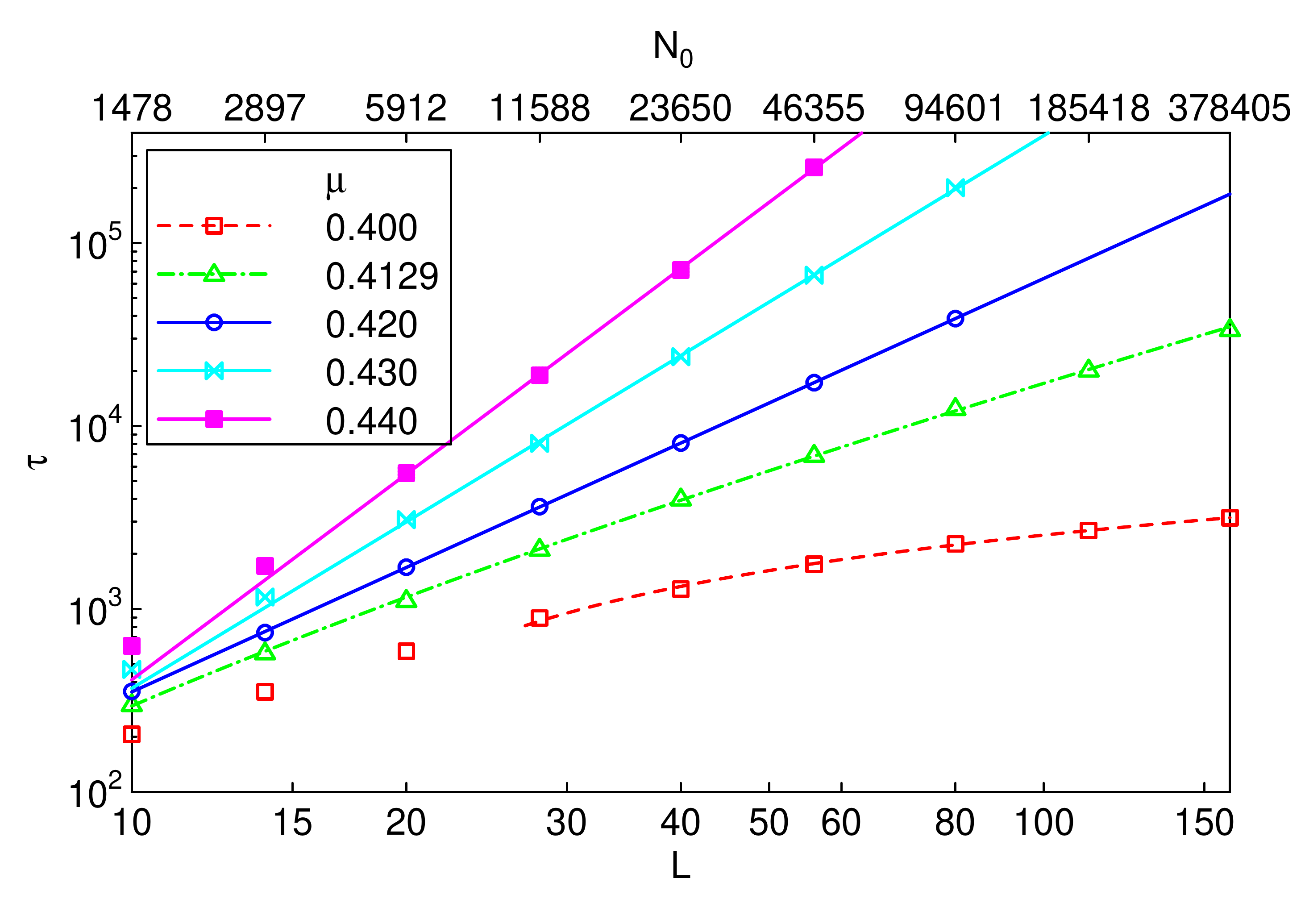}
\caption{Average life time $\tau$ vs.\ system size $L$ (or initial population $N_0$) for several
$\mu$. The solid lines for the $\mu=0.42$, 0.43, and 0.44 curves are power-law fits.
The dash-dotted line for $\mu=\mu_c=0.4129$ represents a fit to $\tau \sim N_0^{1/2} [\ln(N_0)]^{y_\tau}$.
The dashed line for $\mu=0.400$ corresponds to the simple logarithmic behavior expected
on the inactive side of the extinction transition.
}
\label{fig:lifetime}
\end{figure}

In order to address this question in our model, we determine the survival probability $P(t)$ (the
probability that the population has not gone extinct at time $t$) of finite populations close to
the extinction transition. Figure \ref{fig:ptmu0420} shows $P(t)$ for $\mu=0.420$
and several population sizes.
Fitting the survival probability to the exponential $P(t) \sim \exp(-t/\tau)$
yields the life time $\tau$ for each population size. We perform analogous simulations
for a range of mutabilities from $\mu= 0.4$ on the inactive, extinct side of the extinction transition
to $\mu=0.44$ on the active, surviving side.
The resulting graph of life time $\tau$ vs.\ system size $L$ (or, equivalently, initial population $N_0$)
is presented in Fig.\ \ref{fig:lifetime}.
For mutabilities $\mu > \mu_c=0.4129$ (populations on the active side of the extinction transition),
the figure indeed shows a power-law dependence $\tau \sim N_0^{1/\kappa}$ of the life time $\tau$
on the initial population $N_0$. The exponent $\kappa$ is non-universal; power-law fits of
our data give $\kappa=0.53$, 0.66, and 0.89 for mutabilities $\mu=0.440$, 0.430, and 0.420, respectively.
The real-time renormalization group of Ref.\ \cite{VojtaHoyos15} predicts that $\kappa$ increases
as $\mu$ approaches the extinction transition and (in two space dimensions) reaches the value $\kappa_c=2$ right at $\mu_c$.
We also expect a logarithmic correction to the leading power law analogous to those in eqs.\
(\ref{eq:disordered_Ps}), (\ref{eq:disordered_Ns}), and (\ref{eq:disordered_R}).
We therefore fit the lifetime curve for $\mu=\mu_c=0.4129$ to the function
$\tau \sim N_0^{1/2} [\ln(N_0)]^{y_\tau}$ which leads to a high-quality fit with
$y_\tau \approx 2.5$. On the inactive side of the extinction transition, the
theory predicts a slow logarithmic increase, $\tau \sim \ln(N_0)$, of the life time with
population size. This behavior is indeed found for the
$\mu=0.400$ curve in Fig.\  \ref{fig:lifetime}.

\section{Conclusion}\label{sec:conclusions}

In summary, we have studied the extinction transition of a spatially extended biological
population by performing Monte Carlo simulations of an agent-based model. For time-independent
environments, i.e., in the absence of temporal disorder, we have found the extinction transition
to belong to the well-know directed percolation universality class for both the asexual fission
(bacterial splitting) and the assortative mating reproduction schemes.
In the bacterial splitting case, this behavior is expected from the conjecture by Janssen and Grassberger \cite{Janssen81,Grassberger82}, according to which all absorbing state transitions with a scalar order parameter, short-range interactions, and no extra symmetries or conservation laws belong to the directed percolation universality class. The assortative mating case is more interesting because each organism
mates with its nearest neighbor which can, in principle, be arbitrarily far away.
Even though this introduces a long-range interaction in the model, our simulations still yield
directed percolation critical behavior.

The main part of our paper has been devoted to the effects of temporal environmental fluctuations,
i.e., temporal disorder, on populations close to the extinction transition. The question of whether
or not a given universality class is stable against (weak) temporal disorder is addressed by Kinzel's
generalization \cite{Kinzel85} of the Harris criterion (see also Ref.\ \cite{VojtaDickman16} for
version of the criterion that applies to arbitrary spatio-temporal disorder). According to the
criterion, a critical point is stable against temporal disorder if the correlation time exponent
$\nu_\parallel=z\nu_\perp$ fulfills the inequality $z\nu_\perp> 2$. The directed percolation
universality class in two dimensions features a value $z\nu_\perp \approx 1.29$ (see Table \ref{table:exponents_clean}) which violates Kinzel's criterion, predicting that temporal disorder
must qualitatively change the extinction transition.

Simulations of our model in which the death probability $p$ varies randomly from
generation to generation confirm this expectation. They yield an unconventional extinction
transition characterized by logarithmically slow population decay and enormous
fluctuations even for large populations. The time-dependence of the average population size,
the spreading of the population from a single organism, and the mean time to extinction
are well-described by the theory of infinite-noise critical behavior \cite{VojtaHoyos15}.
This is also important from a conceptual statistical mechanics point of view because
it confirms the universality of the infinite-noise scenario by showing that it holds
for off-lattice problems as well as for simple lattice models
such as the contact process with temporal disorder \cite{BarghathiVojtaHoyos16}.

It is worth pointing out that the description of the off-critical behavior
required us to include a correction-to-scaling term. This likely stems from the fact that the
logarithmically slow dynamics leads to a slow crossover to the true asymptotic regime.
This slow crossover may also explain why recent simulations of a similar model
\cite{SKOB17} appear to be compatible with directed percolation behavior even though that model
does contain temporal disorder. For weak temporal disorder, the population is expected to show
directed percolation behavior in a transient time regime before the true asymptotic
critical behavior is reached. As the systems in Ref.\ \cite{SKOB17} are much smaller than
ours (up to 30,000 organisms vs.\ $3.3\times 10^7$), their simulations probably do not reach
the asymptotic regime.

The slow crossover to the asymptotic regime also suggests that the power-law scaling
with nonuniversal exponents observed by Jensen in a directed percolation model with temporal disorder
\cite{Jensen96,Jensen05} does not represent the true asymptotic behavior. Instead, the
nonuniversal power laws hold in a transient time interval before the crossover to the logarithmic infinite-noise  behavior.

The temporal disorder considered in the present paper is \emph{uncorrelated},
i.e., ``white noise''. The effects of long-range noise correlations (leading to ``red noise'')
on population extinction have attracted considerable interest in recent years
(see, e.g., Ref.\ \cite{OvaskainenMeerson10} and references therein).
Our simulations can be easily generalized to this case. From the analogy with long-range
correlated spatial disorder \cite{IbrahimBarghathiVojta14}, we expect that
long-range correlations of the temporal disorder further increase its effects and
lead to a change of the universality class.

In the present paper, we have investigated the extinction transition of populations in
a homogeneous landscape. In the evolutionary context of the model this corresponds
to neutral selection. What about the effects of spatial inhomogeneities? According to the
Harris criterion \cite{Harris74} $d\nu_\perp>2$ (where $d$ is the space dimensionality),
(uncorrelated) spatial disorder is a relevant perturbation and destabilizes the directed percolation
universality class.
Renormalization group calculations \cite{HooyberghsIgloiVanderzande03,HooyberghsIgloiVanderzande04}
and extensive Monte Carlo simulations of the contact process \cite{DickisonVojta05,VojtaFarquharMast09,Vojta12}
have established that the resulting critical point is of exotic infinite-randomness kind.
We expect similar behavior for spatially disordered versions of the present model.

If the system contains (uncorrelated) disorder in both space and time, the generalized Harris criterion governing
the stability of the pure critical behavior reads $(d+z)\nu >2$ \cite{VojtaDickman16,AlonsoMunoz01}.
The directed percolation universality fulfills this criterion in all dimensions suggesting that
such spatio-temporal disorder is an irrelevant perturbation, at least if it is sufficiently weak
(see also discussion in Ref.\ \cite{Hinrichsen00}).

This work was supported by the NSF under Grant Nos.\ DMR-1205803 and DMR-1506152.
We acknowledge valuable discussions with S.\ Bahar.

\bibliographystyle{epj}
\bibliography{../00Bibtex/rareregions}

\end{document}